\input{aipcheck}

\documentclass[
    ,final            
  ]
  {aipproc}

\layoutstyle{6x9}

\begin{document}

\title{Alternatives to Hibernation}

\author{T. Naylor}{
  address={School of Physics, University of Exeter, Stocker Road, Exeter, 
EX4 4QL, U.K.}
}

\begin{abstract}
I outline the evidence pertinent to the connection between the nova 
explosion and mass transfer rates in CVs.
I conclude that there is still insufficient evidence to decide whether or
not such a connection exists.
\end{abstract}

\maketitle


\section{The Problem}

Systems in which classical nova explosions have been observed are structurally 
indistinguishable from other cataclysmic variables (CVs).
They all consist of a low-mass late-type, normally main-sequence star
losing mass via Roche-lobe overflow to a white dwarf.
The magnetic field of the white dwarf can cause differences in the
detail of how the accretion flow reaches the white dwarf.
A strong magnetic field will channel the accretion flow straight onto
the white dwarf (a ``polar'' or AM Her star) whilst a weak field will
allow an accretion disc to form, with intermediate cases occurring when
the inner disc alone is disrupted by the field (an ``intermediate
polar'' or DQ Her star).
However, it should be emphasized that all these sub-classes occur
amongst both classical novae and other CVs, again arguing for a close
relationship between the two.
The obvious conclusion to draw is that all CVs eventually undergo a
nova explosion, and that those we classify as old novae, happen to
have had an explosion in the recent past.

One of the outstanding problems in understanding CVs is the large
range of mass transfer rates they have.
If we concentrate on the non-magnetic systems, where the problem is
best understood, we find a range of about a factor of 100, even if we
restrict ourselves to systems with similar orbital periods.
Part of the non-magnetic CV classification system is based on this
difference, with the high mass transfer rate systems being classified
as UX UMa stars, the intermediate ones as VY Scl and Z Cam stars, and
the low mass transfer rate systems as dwarf novae.

\section{The Solutions}

Shara and collaborators \cite{S89} suggested that nova explosions may
hold the key to explaining the differences in mass transfer rate.
In essence their idea is that the explosion co-incides with a period of high
mass transfer rate, and centuries after the explosion the mass
transfer decreases, and finally ceases altogether.
The detached binary then ``hibernates'', until angular momentum loss
brings the system back into contact, mass transfer begins again,
laying down on the white the material for the next nova explosion.

The attraction of what became known as the ``hibernation scenario'' was 
twofold.
First, at the date it was proposed, it appeared that the observed mass
transfer rates in CVs were much higher than those which the models
required for a nova explosion to occur \cite{S86}.
However, whilst the models of Starrfield, Sparks \& Truran still
require low mass transfer rates \cite{SST00}, the Prialnik \& Kovetz
models \citep[e.g.][]{KP97} do undergo
explosions for mass transfer rates similar to those observed in CVs.
The second attractive feature was that a host of evidence seemed to
imply that old novae continued to fade for at least two hundred years
after outburst, as though their mass transfer rates were declining and
the binary heading towards ``hibernation''.
The evidence for this decline was questioned by myself and others
\cite{NCME}, leading us to suggest that the evidence is perfectly
consistent with the idea that the nova explosion is unconnected with 
the mass transfer rate in the binary \cite{MN95}.
Thus if a system was a dwarf nova before the classical nova outburst,
it will be so again immediately afterwards, and continue to be so for
many nova explosion cycles.
For the purposes of contrast, I will refer to this as the ``constant mass 
transfer model'' or simply CMT.

That two apparently orthogonal theories can be consistent with the
observational evidence is, perhaps, surprising.
In this paper I aim to review the current evidence for links between mass
transfer rate and the time of nova outburst.
I shall begin by discussing the evidence for the systems with long orbital
periods, {\it i.e.} greater than 0.2 days.

\section{Are all novae UX UMa stars?}

A few years after the nova explosion, virtually all classical novae
appear to be high mass transfer rate cataclysmic variables.
In the case of the non-magnetic systems, this means they are UX UMa
stars.
Furthermore, their very similar magnitudes before and after the
explosion \cite{R75}, implies that they were also at high mass transfer rates
before the explosion.
Whilst this may, at first sight, appear to support the idea that there
is a mass transfer cycle, at the peak of which the system explodes as
a classical nova, in fact it probably simply tells us that low mass
transfer rate systems only rarely have nova outbursts.
The reason is that it will take them longer to build up the layer of
material required for the runaway \citep[e.g.][]{I92}.
Further evidence that this is the correct interpretation was provided
by the discovery that Nova Her 1960 (V446 Her) is a dwarf nova \cite{H98}, and
the data of \cite{R75} imply it was probably a dwarf nova before its
nova outburst.

\section{The post nova decline}

In the first hundred years after the nova outburst the system
luminosity declines by around 2 magnitudes.
This was first shown by correlating the age of each nova with its
current brightness \cite{V90}, but later also by following individual
systems \cite{D92}.
Whilst such a decline could be caused by a decline in mass transfer
rate, it could also be due to irradiation of the disc and secondary star
by the white dwarf, which
is hot as a result of the nova explosion, and is cooling.
Support for this idea came first from observations of Nova Cyg 1975
(V1500 Cyg) \cite{SN99}.
The observations show that the inner face of the secondary star is
heated by the white dwarf, and that the degree of heating is declining
with time.
The data can be modeled to deduce the temperature of the white dwarf,
which is found to be falling at the rate expected by theory.
This suggests that it is not actually light from the white dwarf
itself which is responsible for the post-nova decline, but the decline
in flux which is reprocessed by the secondary star and accretion disc.
There are calculations of the reprocessing from the disc alone,
which suggest this is correct \cite{S01}.

If white dwarf cooling is the reason for the post-nova decline, more
observations are explained.
First, there is evidence that the decline in magnitude ceases
after about 100 years, as the white dwarf cooling models predict.
Nova Sge 1783 (WY Sge) is now 200 years old, but the binary still
has the magnitude expected for high mass transfer rate system
\cite{SMN96, SRN97}.
Secondly, old novae sometimes seem to to undergo low amplitude dwarf
nova-like outbursts.
It seems these can be explained by the white dwarf irradiation
maintaining the inner disc in the viscous state, whilst an outer
annulus undergoes dwarf nova outbursts \cite{S00}.
The small region of the disc participating in the outburst explains
its low amplitude.

\section{Conclusions for long period systems}

The real problem here is that the predictions of
hibernation and CMT have converged.
The hibernation model predicts that the decline in mass transfer will 
drive a decline in luminosity after the nova outburst.
Presumably this is already happening in Nova Her 1960, and will happen for
Nova Sge 1783, if we wait long enough.
A CMT model accepts there is a decline in luminosity caused by the falling 
irradiation from the white dwarf, but asserts this stops after about a 
hundred years.
In the CMT view, those old novae which show mass transfer rates below the mean 
are the few low mass transfer rate systems we expect to find.
Thus the discovery by \cite{SD95} that Nova Sco 1860 (T Sco) lies at a 
brightness level suggesting a long period dwarf nova, is consistent with
either model.

\section{Short period systems}

Whatever view one takes of the post-nova decline, until very recently
all the evidence supported the idea that mass transfer rate is broadly
the same before and after the explosion.
Again both theories accept this, as the hibernation models bring the CV up 
to the high mass transfer state before the explosion, and the CMT model 
requires it.
Such a picture, though, was built up when all
old novae with reliably determined orbital periods
had periods longer than about 0.2 days. 
This has now begun to change, and as new systems are being discovered,
there are signs that the picture presented above may not apply to
short period systems.

\begin{figure}
  \includegraphics[height=.3\textheight]{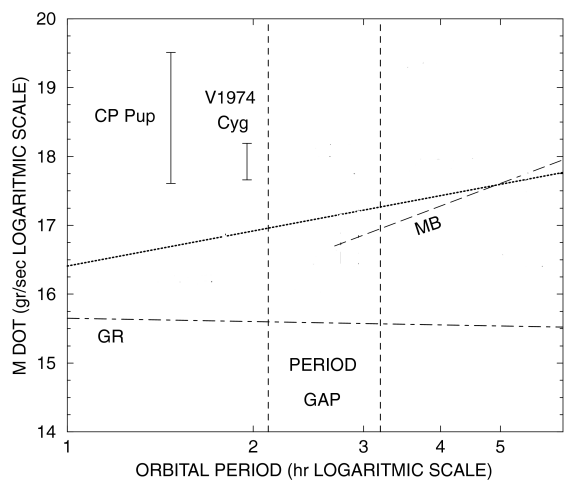}
  \caption{
The orbital period vs mass transfer rate plane for cataclysmic 
variables.  
The two dotted vertical lines mark the approximate limits of the
period gap.
The lines marked MB and GR show the expected mass transfer rates for magnetic
braking and gravitational radiation respectively, and the unlabeled line 
divides high mass accretion rate (stable) discs from low mass accretion rate
ones.
The points for CP Pup and V1794 Cyg are marked.
Adapted from \cite{RN00}.
}
\end{figure}

Perhaps the most important point to make first is that high mass
transfer rate, short orbital period systems simply should not exist.
Figure 1 shows the usual way of classifying non-magnetic CVs, in a
plot of mass transfer rate against period \cite{O96}.
For short periods the only angular momentum loss mechanism
available is gravitational radiation, which cannot support (at least in
the long term) the high mass transfer rates observed in the old novae
CP Pup and V1974 Cyg.
Retter and I \cite{RN00} pointed out that these two systems (which are
the best studied ones below the period gap), appear brighter after
the nova outburst than they were before it -- in contrast to the
behaviour of the long period systems.
Duerbeck \citep{D03} suggests that this effect may be more widespread, 
including systems with periods above the period gap, but less than 0.2 days.

Retter argues that these results suggest that mass transfer rates below the
period gap are driven by the nova explosion.
After the nova explosion the mass transfer rate is high, before dropping
decades or centuries later.
Although this is broadly a ``weak hibernation scenario'', one should note
a crucial difference -- the system is faint immediately before the nova 
outburst, whilst in the normal hibernation models it is bright.
If mass transfer rates do vary in this way, it would explain
why there are some systems well above
the mass transfer rate allowed by gravitational radiation; they are only
there for a short while as a result of the nova event, and will eventually
fall back to their original level.
I have my own reservations.
First, there is no good physical theory of how the cycles would work since
simple mass loss cannot drive them \cite{K01}, nor can irradiation \cite{K96}.
Secondly, due to their small orbital separation, these short period systems 
are the ones we would expect to show the greatest effects of irradiation.
Thus systems which were probably dwarf novae before their nova explosions,
like Nova Cyg 1992 (V1974 Cyg), will find their discs being held in the 
bright state by irradiation for a relatively long time by the mechanism 
outlined in \cite{S00}.
Clearly one solution to this debate is to wait, and see if as the irradiation
declines, V1974 Cyg begins to show dwarf nova outbursts.
However, as the white dwarf takes decades to cool, quicker resolutions would
clearly be preferable.

\section{Conclusions}

For the long orbital period systems ($>$0.2 days), it is clear we see
considerable irradiation from the white dwarf in the decades after outburst, 
which certainly raises the overall luminosity.
Whether when this phase is over, the binary simply returns to its pre-outburst
state, or mass transfer then declines into ``hibernation'' remains an open
question.
For the systems with orbital periods below 0.2 days, there is emerging 
evidence that they are fainter before outburst than afterwards.
This may simply be because most of them are low mass transfer rate systems
(dwarf novae) whose post outburst luminosity is held high for a few decades
by intense irradiation.
Alternatively short period systems may have mass transfer cycles driven by 
the explosion.
Discovering which of these scenarios is correct is crucial to our 
understanding not only of classical novae, but also of CVs in general.


\begin{theacknowledgments}

My thanks are due to Koji Mukai, who not only first kindled my interest
in classical novae, but has also collaborated in developing many of the ideas
presented in our papers reviewed here.
Nye Evans nurtured that interest, whilst Fred Ringwald, Mark Somers and Alon
Retter not only contributed ideas, but did much of the hard work.
Finally, without Mike Shara to argue with, the field would have been much
duller.

\end{theacknowledgments}


\bibliographystyle{aipproc}   

\bibliography{text}

\IfFileExists{\jobname.bbl}{}
 {\typeout{}
  \typeout{******************************************}
  \typeout{** Please run "bibtex \jobname" to optain}
  \typeout{** the bibliography and then re-run LaTeX}
  \typeout{** twice to fix the references!}
  \typeout{******************************************}
  \typeout{}
 }

\end{document}